# Hybridization and localized flat band in the WSe$_2$/MoSe$_2$ heterobilayer grown by molecular beam epitaxy

Lama Khalil[1], Debora Pierucci[1], Emilio Velez[2], José Avila[3], Céline Vergnaud[2], Pavel Dudin[3], Fabrice Oehler[1], Julien Chaste[1], Matthieu Jamet[2], Emmanuel Lhuillier[4], Marco Pala[1], and Abdelkarim Ouerghi[1,∥]

[1] Université Paris-Saclay, CNRS, Centre de Nanosciences et de Nanotechnologies, 91120, Palaiseau,
[2] Université Grenoble Alpes, CEA, CNRS, Grenoble INP, IRIG-Spintec, 38054, Grenoble, France
[3] Synchrotron-SOLEIL, Université Paris-Saclay, Saint-Aubin, BP48, F91192 Gif sur Yvette, France
[4] Sorbonne Université, CNRS, Institut des NanoSciences de Paris, INSP, F-75005 Paris, France

Nearly localized moiré flat bands in momentum space, arising at particular twist angles, are the key to achieve correlated effects in transition-metal dichalcogenides. Here, we use angle-resolved photoemission spectroscopy (ARPES) to visualize the presence of a flat band near the Fermi level of van der Waals (vdW) WSe$_2$/MoSe$_2$ heterobilayer grown by molecular beam epitaxy. This flat band is localized near the K point of the Brillouin zone and has a width of several hundred meVs. By combining ARPES measurements with density functional theory (DFT) calculations, we confirm the coexistence of different domains, namely the reference 2H stacking without layer misorientation and regions with arbitrary twist angles. For the 2H-stacked heterobilayer, our ARPES results show strong interlayer hybridization effects, further confirmed by complementary micro-Raman spectroscopy measurements. The spin-splitting of the valence band at K is determined to be 470 meV. The valence band maximum (VBM) position of the heterobilayer is located at the Γ point. The energy difference between the VBM at Γ and the K point is of -60 meV, which is a stark difference compared to individual 1L WSe$_2$ and 1L WSe$_2$, showing both a VBM at K.

The successful synthesis of graphene by mechanical exfoliation[1] has triggered an intense interest in transition metal dichalcogenides (TMDs; *e.g.*, MX$_2$ with M = W, Mo, etc. and X = S, Se, etc.), as two-dimensional (2D) semiconductors characterized by layer-number dependent optical and electronic properties[2,3]. Van der Waals (vdW) heterostructures[4] combining two monolayers of different 2D-TMD crystals, referred as heterobilayers[5], can further modulate these properties through the choice of materials and the relative interlayer alignment[6,7]. These heterobilayers provide an exciting platform for exploring peculiar physical effects, such as ultrafast charge transfer[8], long-lived spin valley polarization[9,10], exciton condensation[11] and moiré excitons[12–14].

Despite the considerable progress in the understanding of interlayer coupling and the resulting electronic properties[15], there are still many questions to be addressed. Chief among them is the effect of the in-plane twist angle between the two constitutive monolayers (*i.e.*, their relative angular misorientation) on the band structure of the heterojunction. This twist angle governs many electronic features, including band dispersion, electronic bandgap, valence band maximum (VBM) position, excitonic spectrum and spin-orbit coupling (SOC) strength[16]. Most interestingly, quasi-lattice-matched heterobilayers such as 1H-WSe$_2$/1H-MoSe$_2$ exhibit a range of twist angles which gives rise to moiré superlattices[17–20] and host a variety of correlated particle physics[21]. For selected twist angles, the electronic band structure can even exhibit a nearly flat band dispersion[22] with enhanced electron-electron correlations. Although these theoretical prospects are appealing, the experimental exploration TMD-based moiré phenomena is particularly challenging, as it requires both the exact knowledge of the relative heterobilayer misorientation in real space and a fine energy and momentum resolution in reciprocal space.

Recently, a scanning tunneling spectroscopy (STS) study has successfully imaged moiré flat bands in WSe$_2$/WS$_2$ superlattices[23]. Previous scanning tunneling microscopy (STM) studies have also established site-dependent electronic structures in TMD moiré superlattices[20,22,24]. However, the direct visualization of flat bands at the valence band edge of TMD heterobilayer structures has not been demonstrated. Here, angle-resolved photoemission spectroscopy (ARPES) provides a powerful tool with a direct determination of the electronic structure and the related flat band dispersion in moiré superlattices. However, high-resolution ARPES requires a conductive support and thus imposes additional challenges for the material growth, mostly solved by the use of graphene substrates.

Here, we report the direct observation of the moiré flat band electronic structure in 1ML WSe$_2$/1ML MoSe$_2$ vdW heterobilayer (with 1 ML referring to a single monolayer), grown by molecular beam epitaxy (MBE) on graphene substrates. Using ARPES, a flat valence band is observed near the Fermi level and has a width of several hundred meVs. By combining experimental ARPES measurements with density functional theory (DFT) calculations, we show the coexistence of different domains, namely, the 2H stacking and twisted $\theta_{\pm 4}$-domains (with $\theta_{\pm 4} = \pm 4.0°$ twist angle). We attribute the existence of the localized valence flat band to these particular sub-sets of twisted $\theta_{\pm 4}$-domains.

**RESULTS AND DISCUSSION**

We describe here in brief the sample fabrication procedure; more details can be found in the Materials and Methods section. The decomposition of a 6H-SiC(0001) substrate at high temperature under atmospheric pressure in an argon atmosphere was used to produce a graphene monolayer with high electron mobility[25–31] (label "Gr" in Fig. 1 refers to graphene). The graphene monolayer was obtained by tuning the annealing temperature of the initial

SiC substrate[32]. The obtained graphene/SiC substrate was then used to grow 1 ML MoSe$_2$ and 1 ML WSe$_2$ *via* separate but consecutive vdW epitaxy processes[33–37]. Figures 1(a) and 1(b) show side and top schematics of the WSe$_2$/MoSe$_2$ heterobilayer in the 2H stacking, respectively. Each MX$_2$ monolayer contains three atomic layers, X-M-X with the transition metal atoms M covalently bounded to six chalcogen atoms X, forming a sandwiched material in a trigonal prismatic structure (1H phase). In Figure 1(c), we present the reflection high energy electron diffraction (RHEED) pattern of the initial graphene/SiC substrate. We confirm the mono-crystalline property of the graphene layer with in-plane six-fold symmetry. Due to the low incidence of the RHEED electron beam, the illuminated area is several tens of microns wide[38]. The RHEED pattern after the first 1ML MoSe$_2$ growth is shown in Figure 1(d). The graphene-related lines have completely disappeared, which indicates an overwhelming surface coverage by MoSe$_2$. The RHEED pattern consists in regular long streaks, which suggests a well-crystallized order and a flat MoSe$_2$ surface. We find that the hexagonal MoSe$_2$ and graphene unit cells are aligned in-plane ([11-20]$_{MoSe_2}$//[11-20]$_{graphene}$ and [1-100]$_{MoSe_2}$//[1-100]$_{graphene}$). Figure 1(e) shows the RHEED pattern of the 1ML WSe$_2$ layer deposited on MoSe$_2$. Due to the near lattice-match (0.18% difference) between MoSe$_2$ (a = 3.288 Å)[39] and WSe$_2$ (a = 3.282 Å)[39] and their similar hexagonal structure, the RHEED pattern is expected to undergo no visible change from MoSe$_2$ to WSe$_2$. The persistence of the elongated streaks, from the early stage right to the end of the WSe$_2$ growth process, suggests a good vdW epitaxy between the 1ML WSe$_2$ and 1ML MoSe$_2$, with unit cell alignment. This first picture of the respective layer alignment obtained by RHEED is complemented by the following micro-Raman (μ-Raman), X-ray photoemission spectroscopy (XPS) and ARPES measurements.

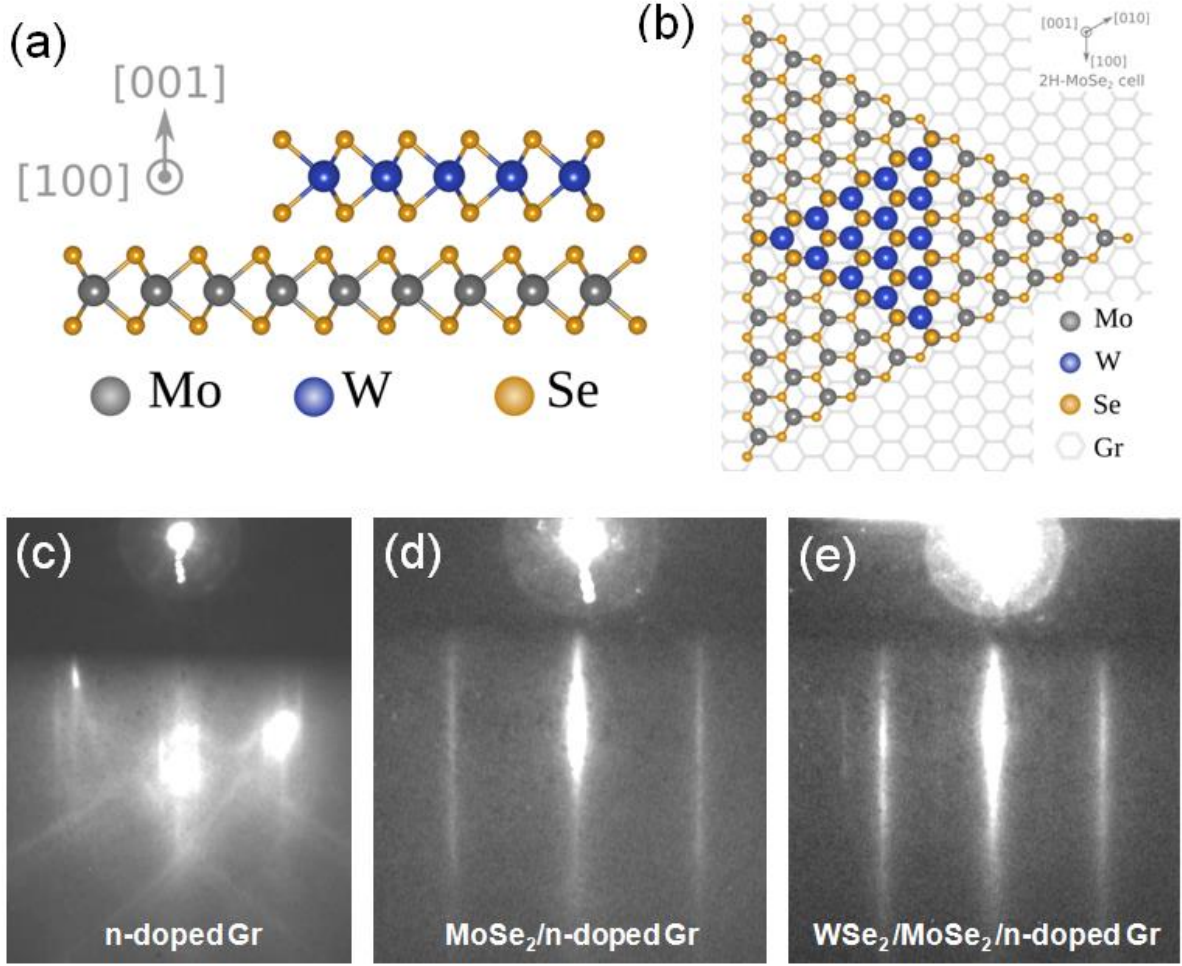

**Figure 1:** (a)-(b) Side and top views of the crystal structure of WSe$_2$/MoSe$_2$ heterobilayer in the 2H stacking. The gray, blue and orange spheres refer to molybdenum, tungsten and selenium atoms, respectively. The underlying graphene layer (Gr) is shown as a honeycomb. (c) RHEED pattern of the graphene/Si(0001) substrate, showing two streaky-line patterns attributed to graphene and to the interface layer. (d)-(e) Streaky RHEED patterns obtained during the MoSe$_2$ and WSe$_2$/MoSe$_2$ MBE growth process, acquired along the same azimuth as (c).

Before exploring the electronic properties of the heterobilayer, we probe its characteristic vibrational modes using μ-Raman spectroscopy. Figure 2(a) shows the μ-Raman scattering spectrum obtained in the frequency range of 190 - 400 cm$^{-1}$ from the WSe$_2$/MoSe$_2$ heterobilayer, which we compare to two other spectra from 1L 1H-MoSe$_2$ and 1L 1H-WSe$_2$, respectively. According to the group theory analysis for monolayer MX$_2$, the irreducible representation decomposition of the vibrational modes can be expressed by $\Gamma = A_{2u}(IR) + A_{1g}(R) + E_{2g}(R) + E_{1g}(IR + R)$[40]. For pristine MoSe$_2$ (blue curve), the out-of-plane mode $A_{1g}$ and the in-plane mode $E_{1g}$ are observed at 240.5 cm$^{-1}$ and 284.4 cm$^{-1}$, respectively. The absence of the Raman peak related to interlayer interactions around 350 cm$^{-1}$ confirms the single monolayer thickness of the reference 1H-MoSe$_2$ film. Similarly, the reference 1H-WSe$_2$ spectrum (red curve) displays the $A_{1g}$ Raman peak at 247.3 cm$^{-1}$ and another mode at about 264.4 cm$^{-1}$, which corresponds to the second order 2LA(M) mode and involves two-phonon scattering close to the M point of the Brillouin zone. The lack of the Raman peak $B^1_{2g}$ at about 300 cm$^{-1}$ is the fingerprint of single 1L 1H-WSe$_2$. Finally, spectrum obtained from the heterobilayer (black curve) only consists of the distinctive above-mentioned $A_{1g}$ (MoSe$_2$), $E_{1g}$ (MoSe$_2$), $E_{1g}$ and $A_{1g}$ (WSe$_2$), and 2LA(M) (WSe$_2$) modes, in agreement the previous report of Nayak *et al.*[41] We note that each of these modes is located at the same vibration frequency as those from the individual 1L references. This confirms that the heterobilayer consists of 1L MoSe$_2$ and 1L WSe$_2$ stacked vertically as a 2H polytype. While we cannot formally exclude the 3R stacking[42,43], we note that 2H (C7 stacking[42,43]) is prevalent for similar WSe$_2$/MoSe$_2$ heterobilayers obtained by MBE growth[42].

To investigate the quality of the WSe$_2$/MoSe$_2$ interface, we then conducted a XPS study using a photon energy of 350 eV. Figure 2(b) shows a wide XPS spectrum, revealing the presence of six core level peaks. Four of these peaks are assigned to the heterobilayer (namely, Mo 3d, Se 3p, Se 3d and W 4f), while the two other peaks are attributed to the graphene/SiC substrate (namely, C 1s and Si 2p). For the Mo 3d, Se 3d and W 4f peaks, we have also recorded high-resolution XPS spectra, as shown in Figure 2(c)-(e), respectively. The experimentally measured data are displayed as black circles and the fitted envelopes are represented as solid blue lines. After a Shirley background subtraction, the core level spectra were fitted by sums of Voigt functions. The fitting of the Mo 3d peak (Figure 2(c)) yields a main doublet (Mo 3d$_{5/2}$ and Mo 3d$_{3/2}$ with a spin orbit (SO) splitting = 3.14 eV) related to stoichiometric MoSe$_2$[44], with the 3d$_{5/2}$ component centered at a binding energy (BE) of 229.1 eV. Beside the Se 3s peaks[45,46] (Se-Mo at BE = 229.8 eV and Se-W at BE = 230.2 eV), we also notice the presence of a second Mo doublet, shifted by 0.45 eV to lower BEs with respect to the main one, corresponding to defective/substoichiometric MoSe$_2$[47]. For the Se 3d core level (Figure 2(d)), we observe two doublets (Se 3d$_{5/2}$ and Se 3d$_{3/2}$ with a SO splitting = 0.86 eV) corresponding to two inequivalent Se species. The first doublet at higher BE of 54.9 eV is relative to Se atoms embedded in the WSe$_2$ monolayer[42,44,48], while the second one at lower BE of 54.4 eV corresponds to Se atoms in the MoSe$_2$ monolayer[42,49–51]. Similarly, to the Mo 3d peak, the W 4f peak (Figure 2(e)) presents two doublets (W 4f$_{7/2}$ and W 4f$_{5/2}$ with a SO splitting = 2.15 eV): the first one is related to stoichiometric WSe$_2$[44] with the W 4f$_{7/2}$ component at a BE of 32.7 eV; the second one at lower BEs (-0.40 eV) corresponds to defective/substoichiometric WSe$_2$[52]. In all spectra, we find no signature of any additional component (*e.g.*, carbon or oxygen bounds)[53–55], indicating that our samples are not contaminated. Moreover, the absence of extra peaks indicates that there is no covalent bond between MoSe$_2$ and WSe$_2$. This is coherent with the expected vdW epitaxy and corroborates the vdW nature of the WSe$_2$/MoSe$_2$ interface. The combination of RHEED, µ-Raman and XPS thus confirms the thickness and chemical composition of our heterobilayer.

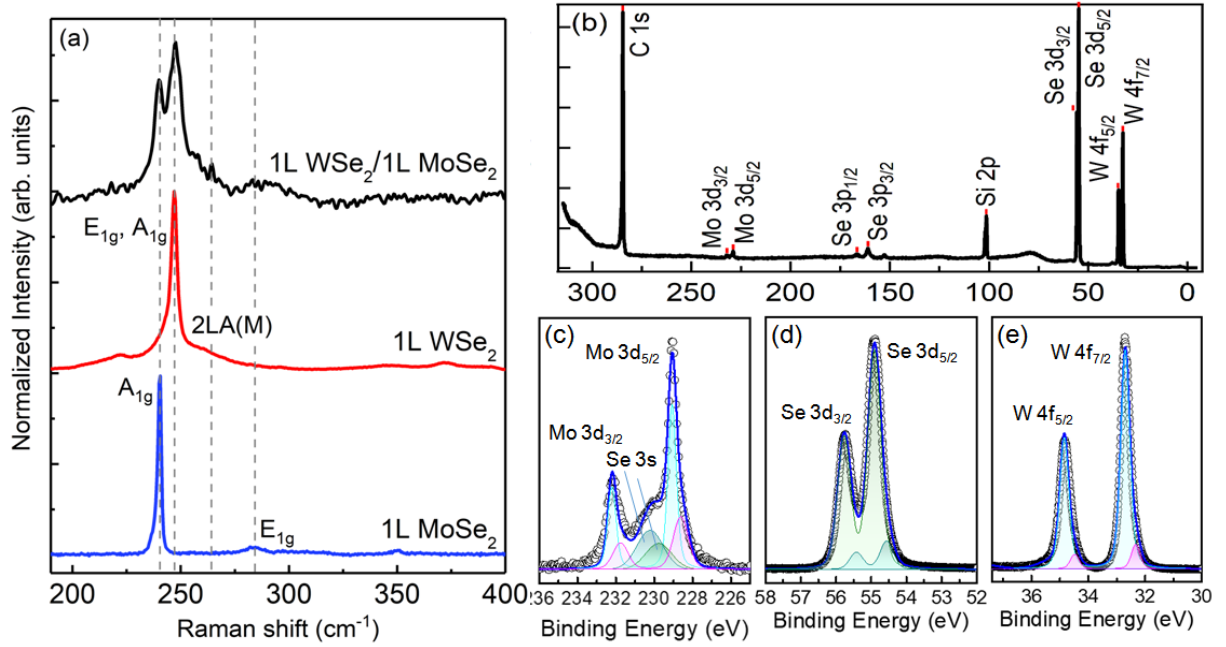

**Figure 2:** (a) Characteristic Raman spectra for 1ML MoSe$_2$, 1ML WSe$_2$ and 1ML WSe$_2$/1ML MoSe$_2$. All measurements were acquired at room temperature using a 532 nm laser excitation with the power of 5 µW (see methods). (b) XPS survey spectrum, acquired with an incident photon energy of 350 eV, revealing all core level peaks related to the heterobilayer structure. (c)-(e) Photoemission spectra of Mo 3d, Se 3d and W 4f, recorded at a photon energy of 350 eV. The experimental data points are displayed as circles, and the solid lines are the envelopes of fitted components.

In the next section, we turn to the local investigation of the electronic band structure of the valence state using nano-ARPES with a small beam size of about 600 nm. Such high spatial resolution will give information relative to the local stacking configuration of our bilayer sample. Figure 3(a) displays a comparison between four valence-band structures along the high symmetry direction ΓK, taken from spatially distinct points on the sample (labelled pt 1 to 4), using an incident photon energy of 100 eV. The sharpness of the different bands is attributed to the high quality of our sample. Near Γ, the bands consist of the W and Mo $d_{z2}$ and Se $p_z$ orbitals[39,56–58]. Because of their out-of-plane character, these bands are the most sensitive to the number of layers composing the system, and thus they might exhibit strong hybridization effects. Close to K, the valence band is spin-split by the intrinsic SOC and the lack of inversion symmetry. The observation of two split branches in the valence band at Γ (marked by orange arrows in Figure 3(a)) for all panels unambiguously confirms the heterobilayer nature of our sample (see band structure calculations for the heterobilayer in figure S1). The same band dispersion and structures are obtained at all probed areas (pt 1 to 4), which supports the spatial homogeneity of the MBE growth.

We remark a lower photoemission intensity near the K point in the nominally gapped region of these electronic spectra, suggesting the existence of localized valence states near the Fermi level. The presence of these localized states can be better visualized by comparing two energy distribution curve (EDC) spectra, Fig. 3(b-c), obtained by integrating the nano-ARPES map (Fig.3(a), pt2) in a wave-vector window of 0.05 Å$^{-1}$ around the Γ point ( k$_∥$ = 0 Å$^{-1}$) and the K point (k$_∥$ = 1 Å$^{-1}$), respectively. In contrast to the EDC at Γ ( Figure 3(b), red), the EDC at K (Figure 3(c), blue) clearly shows a finite density of localized in-gap states near the Fermi level marked as a black arrow, which suggests the presence of a flat valence band state close to the Fermi energy.

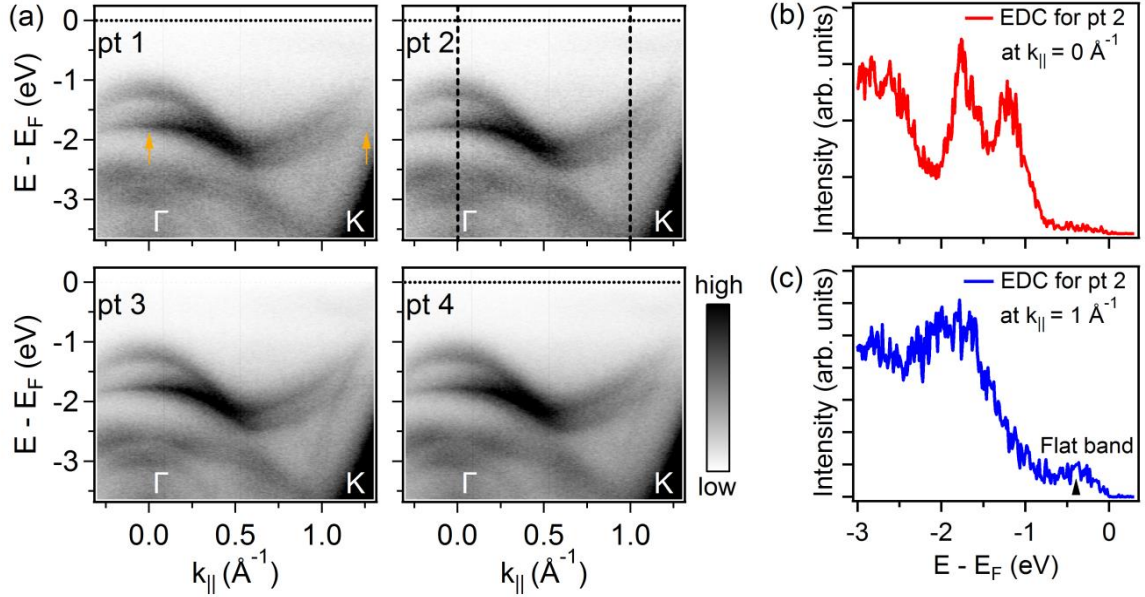

**Figure 3:** (a) Comparison between four nano-ARPES maps of the WSe$_2$/MoSe$_2$ heterobilayer, acquired along the high symmetry direction KΓM from spatially distinct points (labelled pt 1 to 4). The incident photon energy is 100 eV. (b)-(c) Energy distribution curve extracted from the ARPES map (a)[pt2], showing the absence and presence of a flat band close to the Fermi level, near the Γ (K) point, respectively.

Thanks to the spatial homogeneity of the sample (Figure 3), we can better demonstrate the presence of the states near the Femi level in the WSe$_2$/MoSe$_2$ heterobilayer, using a wider 50 × 50 μm$^2$ x-ray beam during the ARPES experiment. By increasing the integration area and the beam flux, we expect to improve the signal to noise ratio and obtain a better visualization of the flat band. This new set of ARPES data was collected with a photon energy of 60 eV. This particular energy value was chosen because the electronic structure of the top-lying valence bands of WSe$_2$ is mostly derived from the W 4d and Se 3p orbitals, each of which possesses a strongly varying photon-energy-dependent photoionization cross section[59]. By lowering the excitation photon energy from 100 eV to 60 eV, the photoionization cross section of these orbitals increases, which directly improves the corresponding ARPES signal[60]. Figure 4(a) presents the photoemission intensity map along the Γ−K direction. In Figure 4(b), we plot EDCs extracted by integrating the spectrum of Figure 4(a) around k$_∥$ = 0 Å$^{-1}$ (Γ, red curve) and k$_∥$ = 1Å$^{-1}$ (K, blue curve). Both the EDC at K (k$_∥$ = 1Å$^{-1}$) and the second-derivative spectrum confirm the presence of a localized in-gap flat valence state (indicated by a black arrow in Figure 4(c)) near the Fermi level.

Compared to nano-ARPES data acquired 100 eV (Figure 3), the conventional ARPES obtained at 60 eV (Fig. 4(a)), shows a better resolution of the valence band near the K point and the states close to the Fermi level. The measured spin-orbit splitting at K is about 470 meV in good agreement with the DFT calculations (see Figure S). The analysis of the k-resolved projected density of states (PDOS) (figure S2) reveals that the highest band is associated to 1ML WSe$_2$, whereas the bands related to 1ML MoSe$_2$ are lower in energy and close to the second band of the WSe$_2$. However, our experimental ARPES data only reveal two bands due to the insufficient energy resolution. For clarity, we report in Figure S3 an ARPES map of a 1ML WSe$_2$ grown by Chemical Vapor

Deposition (CVD) and transferred on a graphene substrate. This monocrystalline CVD WSe$_2$ flake presents a lateral size of about 50-100 µm, which is typically one order of magnitude larger than the typical 30 nm lateral coherence length for the MBE-grown of 2D materials and our WSe$_2$/MoSe$_2$ heterobilayer. Note that 2D materials obtained by epitaxy often presents multiple domains of small lateral dimensions (a few tens nm), in addition to different azimuthal orientations and different thickness. Here, the domain sizes of our WSe$_2$/MoSe$_2$ grown by MBE are well below the resolution of present-generation nano-ARPES instruments, which inevitably results in the spatial averaging of spectra from different domains. This directly affects the collected ARPES data, as we can characterize a single crystal in the case of CVD-grown WSe$_2$, but only a large assembly of MBE-grown WSe$_2$/MoSe$_2$ heterobilayer domains, even with micro- (Figure 4) or nano-sized (Figure 3) X-ray ARPES beams. Such differences in spatial homogeneity and sampling lead to a significant experimental broadening of the top valence bands, from 100 meV for the CVD-grown WSe$_2$ monolayer[61] to over 300 meV for our MBE-grown WSe$_2$/MoSe$_2$ heterobilayer.

We now detail the position of the VBM of the WSe$_2$/MoSe$_2$ heterobilayer. From the ARPES (Figure 4), we determine the VBM at the Γ point, with 1.19 eV BE (binding energy). The energy difference between the VBM and the top of the valence band at the K point (1.25 eV BE) is $\Delta_{\Gamma K} = E_\Gamma - E_K = -60$ meV, which is a stark difference compared to individual 1L WSe$_2$ and 1L MoSe$_2$, showing both a VBM at K[62] (see Supplementary Information Figure S3, comparing ARPES and DFT data on monolayer TMD). Note that such an energy shift does not occur for individual 1ML TMD deposited on graphene, which demonstrates that TMDs and graphene only weakly interact electronically, despite the clean graphene/TMD interface[61]. This VBM shift from K to Γ has been observed in other heterobilayers[62] and is a firm evidence that the two single TMD layers (MoSe$_2$ and WSe$_2$) are strongly coupled electronically. Combined with our previous characterizations, we can draw now a much clearer picture of our sample, which thus consists in a planar stack of three layers (graphene, 1 ML MoSe$_2$ and 1 ML WSe$_2$), with abrupt vdW interfaces, of which the two TMDs (MoSe$_2$ and WSe$_2$) are strongly hybridized, but weakly coupled electronically to the underlying graphene. In a first approximation, our film thus behaves as a simple heterobilayer, independent of the underlying substrate. This has important consequences on the the electronic band structure will strongly depend on the twist angle between the two TMD layers ($\theta_{WSe2/MoSe2}$), but only weakly to angular misalignment with the underlying graphene ($\theta_{Gr/MoSe2}$).

In the following, we will detail the importance of the $\theta_{WSe2/MoSe2}$ value on the band structure of the heterobilayer. In Fig. 4 (c) and (d), we compare the ARPES measurements with density functional theory (DFT) calculations conducted for the 3R and 2H stacked heterobilayer. The DFT calculations (see Methods) are performed over the full k-space (see Supplementary Information, Figure S4). We observe an overall agreement with theoretical DFT calculations computed for 3R and 2H stacking, shown as dashed red and green-colored lines in Fig. 4(c) and (d) respectively. To lift the ambiguity between the two possible 3R and 2H configurations of the 1ML WSe$_2$/1ML MoSe$_2$, we compare their total cohesive energies, defined as the total energy difference per atom between the heterobilayer and the two single monolayers. Our results show that 2H stacking has a lower formation energy than 3R (3meV, see SI (table 1)), so that the cohesive energies are different enough to consider that 2H is the natural stacking configuration in our 1ML WSe$_2$/1ML MoSe$_2$ heterobilayers[63].

We now extract the typical hole effective masses, m*/m$_0$, around the K and Γ points, which are of 0.4 ± 0.1 and 1.2 ± 0.1, respectively. The spin-splitting of the valence band at K is determined to be 470 meV. Combining ARPES and DFT, we expect our heterobilayer to show an indirect band gap (1.31 eV) between the VBM at Γ and the conduction band at K (see green dashed arrow in Fig 4(d)). While there is a relatively good agreement between the DFT and ARPES, there are also important differences. The most prominent one is a flat band at E - $E_F$ = -0.5 eV (with ~390 meV bandwidth), inside the forbidden energy region, near the Fermi level, see Fig. 4(c). Other bands, located between 2.5 eV and 3 eV BE, are also not reproduced by the calculations. Therefore, it is likely that the valence band dispersion in our heterobilayer is modulated by the change of the twist angle between the two layers.

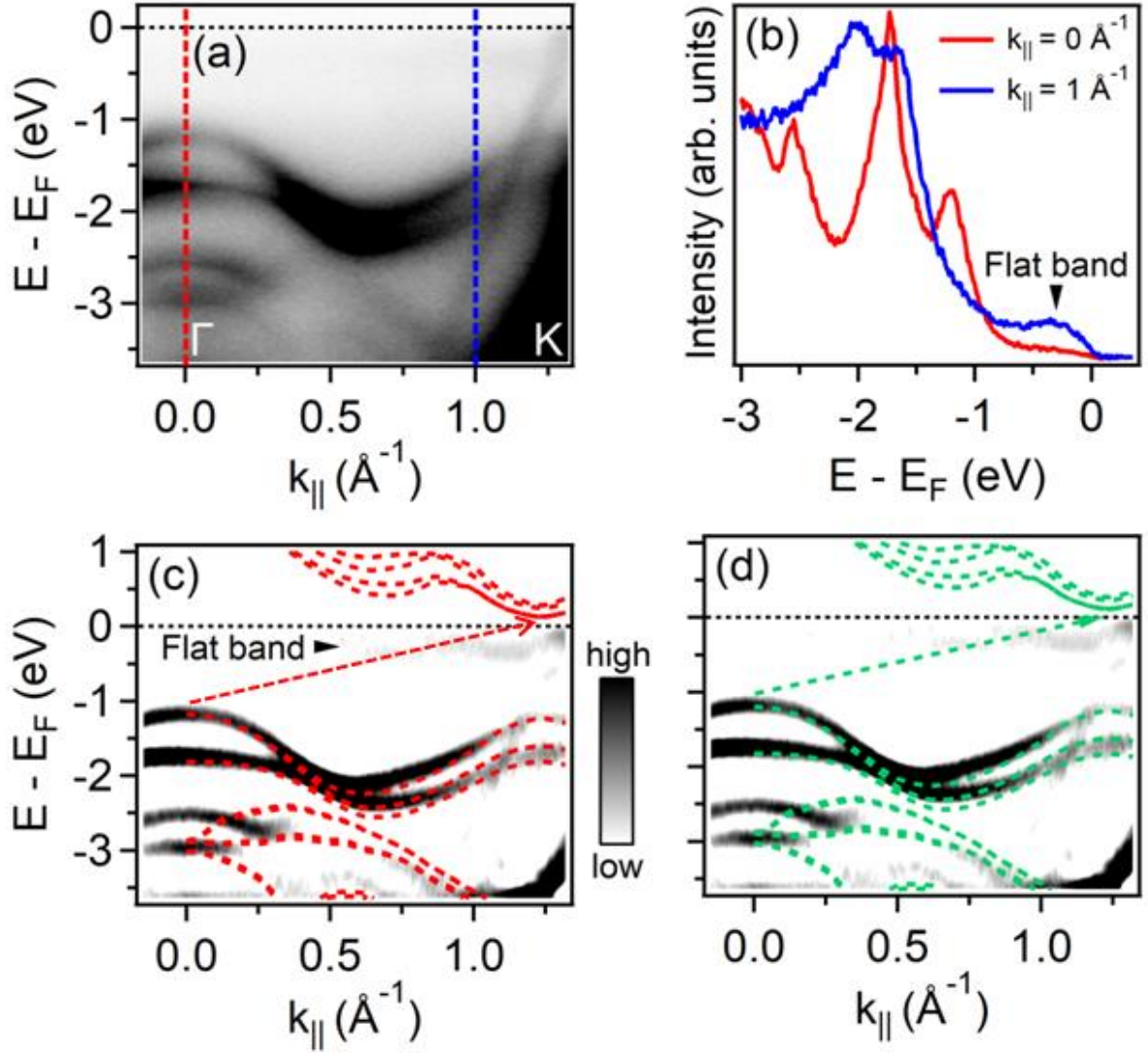

**Figure 4:** (a) High-resolution ARPES intensity map for 1 ML $WSe_2$/1 ML $MoSe_2$, acquired along the KΓM high symmetry direction with a photon energy of 60 eV. (b) Energy distribution curves obtained at $k_\parallel$ = 0 Å$^{-1}$ and $k_\parallel$ = 1Å$^{-1}$ from the ARPES spectrum in (a), confirming the presence of a flat band around the K point, as observed by nano-ARPES. (c and d) Second derivative spectrum obtained from the intensity map in (a), showing the presence of flat bands near the Fermi level. (d) Comparison between experimental ARPES data and theoretical DFT calculations computed for 3R and 2H stacking of 1 ML $WSe_2$/1 ML $MoSe_2$, shown as dashed red and green-colored lines, respectively. The indirect band gap transition is marked as dashed arrow.

To gain more insight into the impact of the twist angle on heterobilayer band dispersion, we acquired three constant ARPES energy maps in the $k_x$-$k_y$ momentum plane, using incident photon energy of 100 eV. The first one showing the graphene Dirac cone at the $K_{Gr}$ point; the second and third ones are obtained at the valence band maxima at the $\Gamma$ and $K_{heterobilayer}$ points, respectively. We can also notice the presence of faint circular features surrounding the horseshoe arc of graphene. These features can be attributed to the π-band replicas of graphene[64]. For perfect 2H stacking, $\Gamma$, $K_{heterobilayer}$ and $K_{Gr}$ are collinear and the relative angles between graphene, MoSe$_2$ and WSe$_2$ are all 0°. Instead of this perfect alignment, we observe an "arc" of the heterobilayer valence band at $K_{heterobilayer}$ (the red arc drawn in Fig. 5(c) is a guide to the eye), which indicates that the TMD heterobilayer presents a range of rotational misalignment in the [-4°, +4°] interval [34]. Such a continuous arc results from the incoherent emission from a large population of separate domains; each arbitrary rotated within the twist angle interval; sampled together by the relatively large ARPES beam size (50 μm). The presence of arbitrarily rotated domains is a typical feature of the weak vdW bonding between the components of the heterobilayer.

Considering that our sample hosts a significant population of misaligned layers with respect to the 2H stacking, it is not surprising to observe discrepancies between the experimental band structure and the DFT simulation. In TMD heterostructures assembled from closely related twisted heterobilayer, MX$_2$/MX$_2$ with a small difference in lattice constant or orientation, theoretical investigations show that the band structure strongly depends on the intralayer twist angle[65–68]. Unlike graphene bilayer with precise 'magic' angles[69], the valence flat bands theoretically predicted in TMD heterobilayers results from hole localization at the moiré high symmetry points due to moiré-induced band gap fluctuations[67]. Assuming the distance between these high symmetry moiré stacking is large enough compared to Coulomb interactions (i.e. low twist angle), the holes will localize due to the band gap fluctuations in spatially separated pockets and behave as localized independent defects. Consequently, their corresponding energy dispersion shapes as a flat band in reciprocal space. In addition, there is no specific twist angle value associated to this type of flat band but rather a limited continuous interval centered at $\theta = 0°$. In the case of 2H stacked WSe$_2$/MoSe$_2$, LDA calculation leads to a moiré potential with amplitude about 10 meV for valence band states associated with WSe$_2$. For this system, the highest valence moiré band is flat and energetically isolated from other bands when the twist angle θ is less than 5.6°. Figure 5(d) shows an example of a moiré configuration for $\theta_{WSe2/MoSe2} = 2.6°$ ($\theta_{WSe2/MoSe2} = 0°$ being the conventional 2H stacking). The resulting hexagonal moiré cell (solid black line) hosts a variety of local stacking, including the conventional 2H at position B but also other high symmetry stacking at positions A and C. Note that small positive or negative twist angle variations (a few degrees) around $\theta_{WSe2/MoSe2} = 0°$ only affect the size of the Moiré cell but not the type of high symmetry stackings. The theoretical twisted angle values leading to a flat band scenario in MoSe$_2$/WSe$_2$[68] are $\theta = 1.6$ and 2.6°, which falls in our experimental interval for WSe$_2$/MoSe$_2$, $\theta_{\pm 4} = \pm 4.0°$, as determined by ARPES. Still, the experimental bandwidth of the flat band is about 390 meV, which is much higher than the expected bandwidth predicted by the theory (between a few tens of meV to a couple of hundred meV)[68,70]. This experimental bandwidth value can be attributed to the simultaneous sampling of multiple domains with small lateral dimensions with respect to the spatial resolution of present-generation nano-ARPES instruments, (see the discussion of Figure 3 and 4), which experimentally results in the broadening of all the valence bands. Additionally, the superposition of signals from domain with different $\theta_{WSe2/MoSe2}$ misorientation angles, each with a slightly different theoretical flat band energy position respect to the Fermi level from DFT results[68,70], also mathematically results in the broadening of the flat band alone.

Conversely, in the case of a specific "magic" angle flat band scenario, the in-plane mosaic spread combined with the relatively large lateral probe size of the ARPES beam would lead a small signal to noise ratio for the ARPES flat band signal, proportional to the low probability of sampling this particular "magic" twist angle value in the total population. In contrast, the flat band scenario described in Ref[67,68] is more robust, as it can occur in a range of twist angle, and thus be captured at much higher signal to noise ratio in our mosaic sample. We therefore conclude that it is very likely that the flat band dispersion observed in our $WSe_2/MoSe_2$ heterobilayer originates from localized states created by the moiré pattern and the corresponding twist angle between the two TMD lattices.

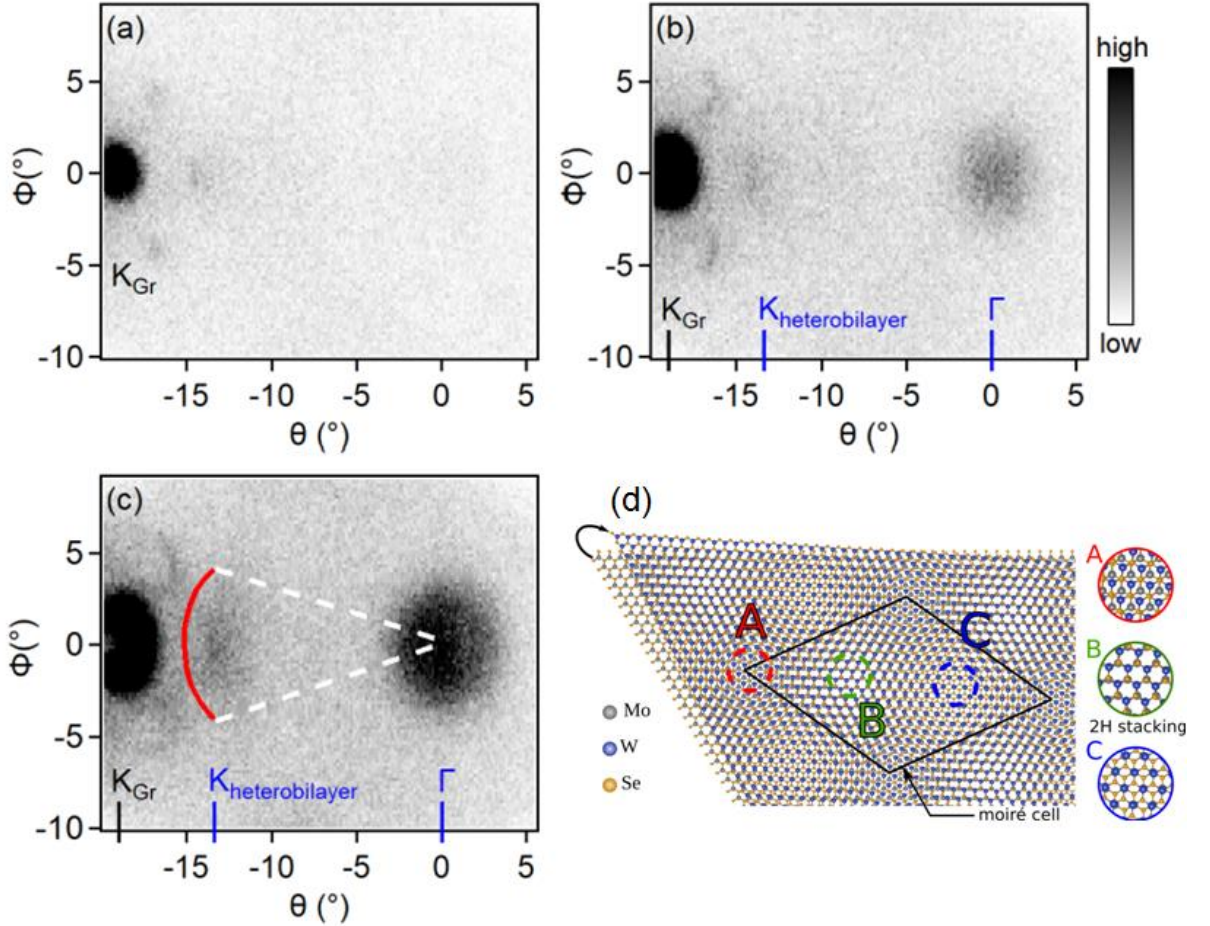

**Figure 5:** Constant energy maps in the ($k_x$, $k_y$) momentum plane, acquired with an incident photon energy of 100 eV. (a) is acquired in the vicinity of the Dirac point of graphene. (b-c) are obtained around the valence maxima at $\Gamma$ and $K_{heterobilayer}$ of the heterobilayer, respectively. (d) Illustration of local moiré stackings (A, B, C) in the moiré cell (solid black line) of $WSe_2/MoSe_2$ with twist angle $\theta_{WSe2/MoSe2} = 2.6°$.

**CONCLUSION**

In summary, we have demonstrated the direct MBE growth of 1L $WSe_2$/1L $MoSe_2$ heterobilayer on graphene *via* a van de Waals epitaxy process. The high-resolution XPS measurements showed the absence of any contamination and revealed sharp and abrupt van de Waals interface between all materials. Our study of the electronic band structure by ARPES revealed an in-plane mosaic spread with twist angles as large as ±4°. A detailed analysis shows that the heterobilayer is strongly coupled electronically. We note the existence of a flat band, positioned inside the bandgap in the vicinity of the Fermi level, which should not exist for perfectly stacked layers. The comparison with theoretical work gives compelling evidence that this flat band relates to twist-dependent moiré superlattices between the two TMD layers. This observation of such flat valence band opens a promising route to engineer the band structure of similar twisted heterobilayer systems.

**ACKNOWLEDGMENTS**

We acknowledge the financial support by RhomboG (ANR-17-CE24-0030), MagicValley (ANR-18-CE24-0007) and Graskop (ANR-19-CE09-0026) grants. This work is supported by a public grant overseen by the French National Research Agency (ANR) as part of the "Investissements d'Avenir" program (Labex NanoSaclay, ANR-10-LABX0035).

**Competing financial interests:** The authors declare that there are no competing interests.

**Data availability:** The datasets generated during and/or analyzed during the current study are available from the corresponding author on reasonable request.

**Contributions:** E.V., M.J. and C.V. fabricated the samples. L.K., D.P., A.O., J.A., and P.D. carried out the nano-XPS/ARPES experiments. J.C., characterized the samples by means of µ-Raman spectroscopy. M.P., carried the DFT calculations. All authors analyzed the results and contributed to the scientific discussions and manuscript preparation.

**MATERIALS AND METHODS**

**Growth of the 1L WSe$_2$/1L MoSe$_2$ heterobilayer:** The heterobilayer was grown by molecular beam epitaxy in a ultrahigh vacuum (UHV) reactor with a base pressure in the low $10^{-10}$ mbar range. Before the growth, the graphene substrate was thermally cleaned in the UHV chamber at a temperature of 800°C for 30 minutes. The heterobilayer growth temperature was 470°C. The direct Se flux measured at the sample position using a retractable gauge was $10^{-6}$ mbar. The Mo/W deposition rate, measured with a quartz balance, was 1.1 Å/min corresponding to the growth of one WSe$_2$/MoSe$_2$ layer in 2 minutes and 30 seconds. Each monolayer was annealed at 780°C under Se during 15 minutes to improve crystalline quality. Finally, the heterobilayer was capped with an amorphous Se layer deposited at room temperature to prevent any degradation of the film during air transfer.

**µ-Raman measurements:** The µ-PL measurements were conducted at room temperature with a x100 objective and a 532 nm laser excitation, using a commercial confocal Horiba micro-Raman microscope. The laser beam was focused onto a small spot having a diameter of ~1 µm on the sample and its incident power was about 5 µW. In our configuration, this laser power is sufficiently low to avoid any shift of the Raman modes on all samples.

**DFT calculations**: Bandstructure simulations of the 2H stacked WSe$_2$-MoSe$_2$ vdW heterobilayer were obtained by performing DFT simulations via the Quantum ESPRESSO suite[69], including spin-orbit coupling and vdW forces. Relevant parameters and models are detailed in the Supplementary Information.

**Band structure of the WSe$_2$/MoSe$_2$ heterobilayer:** The ARPES experiments were conducted at the ANTARES beamline of the SOLEIL synchrotron light source (Saint-Aubin, France). We used linearly horizontal polarized photons of 60 eV, 100 eV and 350 eV and a hemispherical electron analyzer with vertically-confining entrance slit to allow band mapping. The total angle and energy resolutions were 0.25° and 10 meV. All XPS and ARPES measurements were performed at 70 K.

# Hybridization and localized flat band in the WSe$_2$/MoSe$_2$ heterobilayer grown by molecular beam epitaxy

Lama Khalil[1], Debora Pierucci[1], Emilio Velez[2], José Avila[3], Céline Vergnaud[2], Pavel Dudin[3], Fabrice Oehler[1], Julien Chaste[1], Matthieu Jamet[2], Emmanuel Lhuillier[4], Marco Pala[1], and Abdelkarim Ouerghi[1,||]

[1] Université Paris-Saclay, CNRS, Centre de Nanosciences et de Nanotechnologies, 91120, Palaiseau,
[2] Université Grenoble Alpes, CEA, CNRS, Grenoble INP, IRIG-Spintec, 38054, Grenoble, France
[3] Synchrotron-SOLEIL, Université Paris-Saclay, Saint-Aubin, BP48, F91192 Gif sur Yvette, France
[4] Sorbonne Université, CNRS, Institut des NanoSciences de Paris, INSP, F-75005 Paris, France

[||] Corresponding authors E-mail:
abdelkarim.ouerghi@c2n.upsaclay.fr

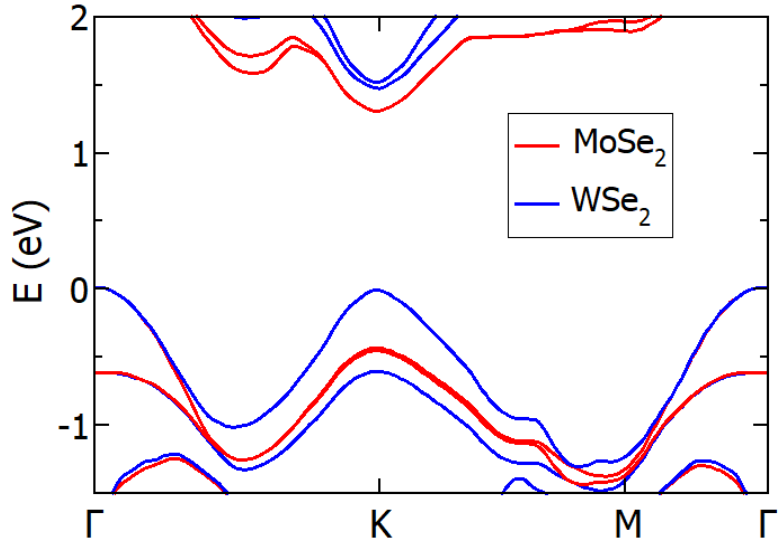

**Figure S1:** Bandstructure simulations of the 2H stacked WSe$_2$-MoSe$_2$ vdW heterobilayer. Electronic properties of the systems were obtained by performing density functional theory (DFT) simulations via the Quantum ESPRESSO [1] suite. Blue and red lines refer to the WSe$_2$ and MoSe$_2$ layers, respectively. The spin-orbit coupling was included by adopting fully relativistic pseudopotentials and non-collinear effects. In order to better estimate the electronic band-gap, we used the HSE hybrid functional [2], as well as a 15x10x1 Monkhorst-Pack grid for the k-vectors and a cutoff energy of 50 Ry. Van der Waals forces were simulated by means of the Grimme-d3 model [3] and a vacuum space of 25 Å was considered along the z axis to minimize the interaction between periodically repeated layers. In order to account for the interaction with the graphene substrate and match the ARPES measurements, we used a lattice parameter of $a$ = 3.3762 Å, thus considering a tensile strain of about 1.8% with respect to the relaxed lattice. Atomic positions were determined after relaxation with convergence threshold for forces and energy of $10^{-3}$ and $10^{-4}$ (a.u.), respectively.

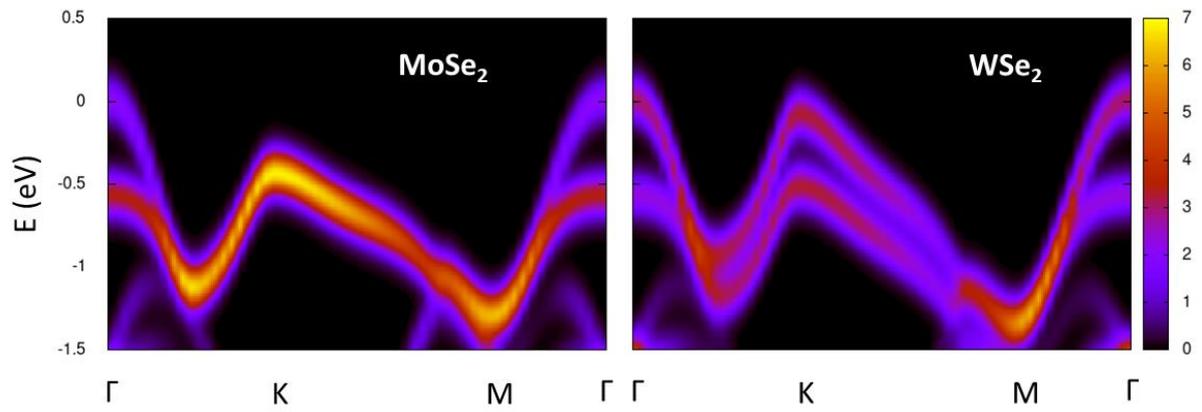

**Figure S2**: k-resolved projected density of states (PDOS) of the $WSe_2$-$MoSe_2$ heterobilayer, showing (left) the PDOS of the $MoSe_2$ layer and (right) the PDOS of the $WSe_2$ layer. The simulation parameters are the same as in Fig. S1.

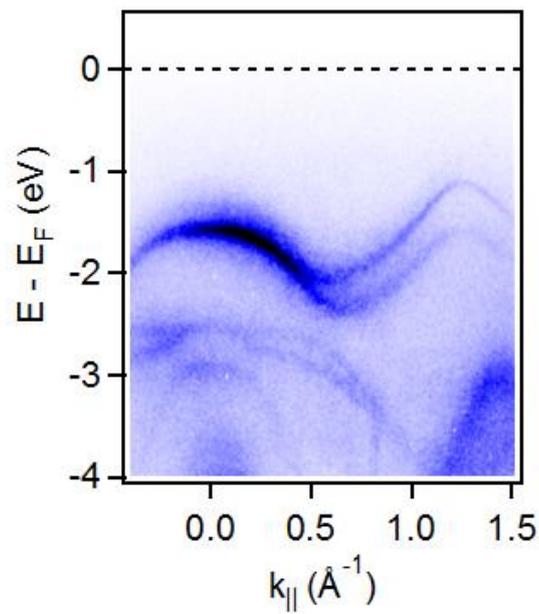

**Figure S3:** $WSe_2$/Graphene heterostructure: ARPES map of single layer CVD-grown $WSe_2$ along the ΓK high symmetry directions at 100 eV.

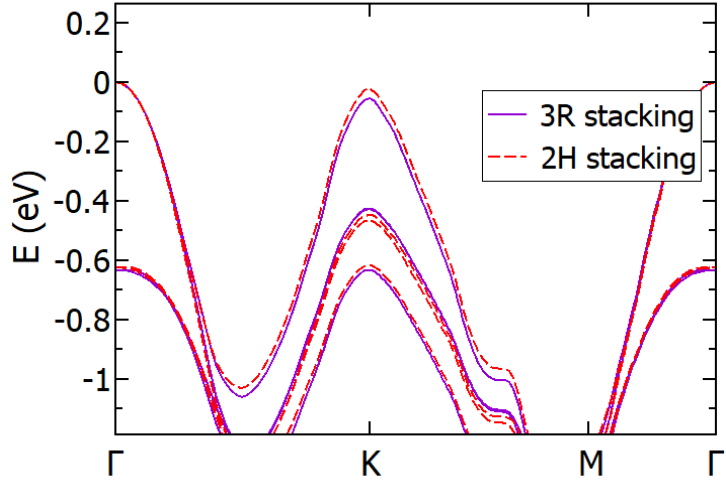

**Figure S4**: Comparison of 3R and 2H stacking of the hetero-bilayer composed of 1ML $WSe_2$ and 1 ML $MoSe_2$

In order to lift the ambiguity between 3R and 2H stackings, we performed additional DFT calculation concerning the total binding energy of the $WSe_2/MoSe_2$ in the 3R and 2H configurations, the results are reported in the following table:

Table 1:

| E($MoSe_2$) (Ry) | E($WSe_2$) (Ry) | E($WSe_2$)+E($MoSe_2$) (Ry) | E(3R) (Ry) | E(2H) (Ry) | CE(3R) (eV) | CE(2H) (eV) |
|---|---|---|---|---|---|---|
| -175.152408 | -192.51491 | -367.667321 | -367.6828 | -367.68309 | -0.2113 | -0.2145 |

Where the cohesive energy of the system is defined as CE ($MoSe_2$–$MoSe_2$) = E ($MoSe_2$-$WSe_2$) - E ($MoSe_2$) - E ($WSe_2$) As we can see the cohesive energy of the 2H is lower than the 3R configuration by 3 meV, indicating a higher stability for this configuration.

**References :**

1. Giannozzi, P. et al. QUANTUM ESPRESSO: A modular and open-source software project for quantum simulations of materials. J. Phys. Condens. Matter 21, 395502 (2009), doi: 10.1088/0953-8984/21/39/395502

2. Heyd, J., Scuseria, G. E. and Ernzerhof, M. Hybrid functionals based on a screened Coulomb potential. J. Chem. Phys. 118, 8207 (2003), doi: 10.1063/1.1564060

3. Grimme S., Semiempirical GGA-type density functional constructed with a long-range dispersion correction. J. Chem. Phys 132, 154104 (2010), doi: 10.1002/jcc.20495